\documentstyle[12pt]{article}
\def\reference{\parskip 0pt\par\noindent\hangindent 0.5 truecm}

\def\hi{H\,{\sc I}}
\def\xp#1{{10}$^{{#1}}$}

\def\cmc{cm$^{-3}$}
\def\xxp#1{$\times${10}$^{{#1}}$}

\input{psfig.sty}
\begin{document}
%
%
\title{HI Absorption in GPS/CSS Sources}
%


\author{Ylva Pihlstr\"om $^{1}$,
John Conway $^{2}$ \& Rene Vermeulen $^{3}$
} 

\date{}
\maketitle

{\center
$^1$ NRAO, P.O. Box O, Socorro, NM 87801, USA\\ypihlstr@nrao.edu\\[3mm]
$^2$ Onsala Space Observatory, S-439 92 Onsala, Sweden\\jconway@oso.chalmers.se\\[3mm]
$^3$ ASTRON, Postbus 2, 7990 AA Dwingeloo, The Netherlands\\rvermeulen@astron.nl\\[3mm]
}

%

\begin{abstract}
Combining our own observations with data from the literature, we
consider the incidence of \hi\ absorption in Gigahertz Peaked Spectrum
(GPS) and Compact Steep Spectrum (CSS) sources. Here we present our
preliminary results, where we find that the smaller GPS sources ($<$~1
kpc) on average have larger \hi\ column densities than the larger CSS
sources ($>$~1 kpc). Both a spherical and an axi-symmetric gas
distribution, with a radial power law density profile, can be used to
explain this anti-correlation between projected linear size and \hi\
column density. Since most detections occur in galaxy classified
objects, we argue that if the unified schemes apply to the GPS/CSSs, a
disk distribution for the \hi\ is more likely.
\end{abstract}

{\bf Keywords: galaxies: active -- galaxies: ISM -- radio lines: galaxies}

\bigskip

%
%

\section{Probing the gas content in GPS/CSS sources}
Attempts to study the internal gas properties of GPS/CSS sources have
been made in several different wavebands. For instance, optical
studies show strong highly excited line emission with large equivalent
widths (Gelderman \& Whittle 1994), consistent with interactions
between the radio source and the interstellar medium. The majority of
the small GPS sources are weakly polarised (Stanghellini et al.\
1998), suggesting strong depolarisation consistent with a very large
central density. Other evidence for a dense environment comes from
free-free absorption observations of for instance OQ208 (Kameno et
al.\ 2000) and NGC1052
(e.g.\, Kellermann et al.\ 1999; Vermeulen et al.\ 2002a,
2002b). Another example is Marr et al.\ (2001) who found evidence for
free-free absorption in the GPS 0108+388, consistent with a 100~pc
radius disk with an electron density of 500~\cmc. Also in 1946+708
multi-frequency continuum studies show indications of free-free
absorption concentrated toward the core and inner parts of the
counter-jet, again suggesting a disk or torus origin (Peck et al.\
1999). Disk-like kpc-scale distributions of gas have also been found
in the optical, and the best example so far is the HST dust disk
observed in the GPS 4C\,31.04 (Perlman et al.\ 2001).

Another way to study the gas content in these sources is by spectral
absorption experiments, which are advantageous mainly because the
sensitivity is independent of the source redshift. With this method
small masses of gas may be probed, although high column densities are
required. Using the 21 cm line of atomic hydrogen in absorption it is
possible to study the atomic hydrogen (\hi) content of GPS/CSS
sources. The \hi\ is likely only a fraction of the total gas present,
therefore the observations provide lower limits to the total gas mass
and density. The strength of GPS/CSS sources at cm wavelengths makes
them good targets for such experiments, in addition their small sizes
indicate that lines of sight to CSSs will sample the dense gas
confined within the centre of the host galaxy. Similarly the line of
sight to a GPS source will trace gas within the narrow-line region
(NLR).

\section{The sample}
In order to increase the statistics of high redshift sources with
associated \hi\ absorption, a survey to detect redshifted \hi\ in
northern sky sources has been performed using the Westerbork Synthesis
Radio Telescope (WSRT). The WSRT is equipped with wide bandwidth UHF
receivers (700-1200 MHz), which enables studies of the redshifted
$\lambda =$~21 cm line of neutral hydrogen for 0.19~$< z<$~1.0. Around
60 GPS/CSS sources have been searched at the WSRT so far, as parts of
several different projects with slightly varying goals (Vermeulen et
al.\ 2002, in prep.). From those sources we select objects which have
projected linear sizes $<10$ kpc, and in addition to our own
observations we have included lower redshift GPS/CSSs from the
literature which had available HI absorption data (see Pihlstr\"om
2001).

\section{Relation between N$_{\rm HI}$ and linear size}
Including all targets, our preliminary results show a general \hi\
absorption detection rate of 48\%. However, if we make the GPS/CSS
division at 1~kpc, we find that the GPS sources have a detection rate
of 53\% as compared to the CSS detection rate of 36\%. This could
reflect the fact that the more compact sources have a larger part of
their continuum emission covered by nuclear gas. This effect has
already been suspected due to the high detection rate of \hi\
absorption in nearby GPS/CSS objects (Conway 1996).

The probable youth of GPS/CSSs implies the possibility of studying the
birth of radio sources, and little is known of the mechanisms
triggering the radio activity. It has been suggested that mergers
could provide a way to transport gas to the centre of the host
galaxies and thus be involved in the onset of the nuclear
engine. Indeed, optical observations have shown that many of the
GPS/CSS sources are in disturbed or interacting systems (de Vries et
al.\ 2000; O'Dea et al.\ 1996). If the GPS/CSS sources are young
sources resulting from mergers, we expect a gas rich galactic
nucleus. The amount of gas needed to fuel the AGN is probably much
smaller than the total gas mass available, thus we do not expect the
total gas mass to be systematically different between sources which
are all younger than a million years. However, a radial density
profile may be reasonable to assume, and then absorption experiments
would probe different column densities of gas as the source grows in
size (since most of those objects are lobe-dominated). Using results
from our \hi\, absorption studies together with results compiled from
the literature, we here investigate if the amount of neutral atomic
gas shows any systematic variation with source size.

\begin{figure}[t]
\begin{center}
\hspace*{0.0cm}\psfig{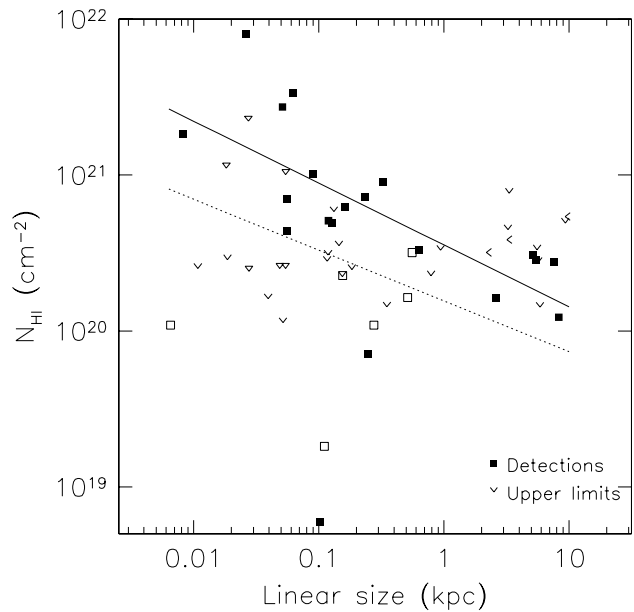}
\caption{Absorbed \hi\ column density versus projected linear
size. Square symbols are detections, while the arrows denote upper
limits at a 3$\sigma$ level. Arrows pointing to the left are points
for which we only have upper limits on the source size. Open squares
and open triangles indicate preliminary data still being
reduced. There is an anti-correlation between the source size and the
amount of absorbing gas, confirmed by survival analysis. The least
squares fit taking into account the upper limits is shown with the
dotted line. This can be compared with a least square fit to the
detections only, plotted with the solid line. }
\label{nhi_size}
\end{center}
\end{figure}

Despite that a few numbers are still tentative (data reduction is
still in progress for a number of objects), there appears to be an
anti-correlation between the source linear size and the \hi\ column
density (Fig.~\ref{nhi_size}). A visual inspection suggests that the
column densities for sources $<$~1 kpc are larger than for those
$>$~1kpc. The data set contains upper limits both in \hi\ column
density, as well as a few points which have upper limits on their
size. We therefore investigated the possible slope in
Fig.~\ref{nhi_size} using the survival analysis package ASURV, taking
into account also the upper limits (Lavalley et al.\ 1992). A
Kendall's Tau test show that there exists a correlation between the
column density and the linear size with a probability $>$~99\%, while
a Spearman's Rho test shows a correlation with a probability
$>$~95\%. Linear regression applied with ASURV finds the relationship
$N_{\rm HI}=$\xp{20.2}$ LS^{-0.33}$; where $LS$ is the linear size in
kpc. This linear fit is plotted with a dotted line in
Fig.~\ref{nhi_size}.

An interesting question is whether the population of small sources
have larger column densities because the FWHM of the absorption line
is wider, or because the line is deeper. Spearman rank tests show no
significant correlation between the observed column densities and line
widths, while the peak optical depths and linear sizes are correlated
with probability $>$95\%. This is shown in Figure 2, where the solid
line represents a least square linear fit to the data points. The
slope is $-$0.32, which implies that the slope in Fig.~\ref{nhi_size}
mainly depends on a difference in the observed opacity and is not due
to differences in the line width.

\begin{figure}[t]
\begin{center}
\hspace*{0.0cm}\psfig{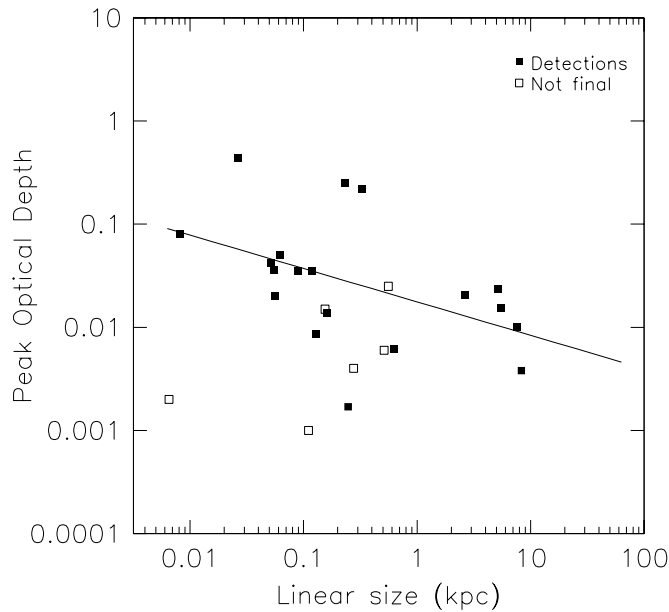}
\caption{Peak optical depth versus linear size. Filled squares
represent \hi\ absorption detections, while open squares are
preliminary detections. A clear correlation is found with a
probability $>$~95\%, which is enough to explain the
correlation in Fig.\ \ref{nhi_size}.}
\label{tau_size}
\end{center}
\end{figure}

\begin{figure}[ht]
\begin{center}
\hspace*{0.0cm}\psfig{file=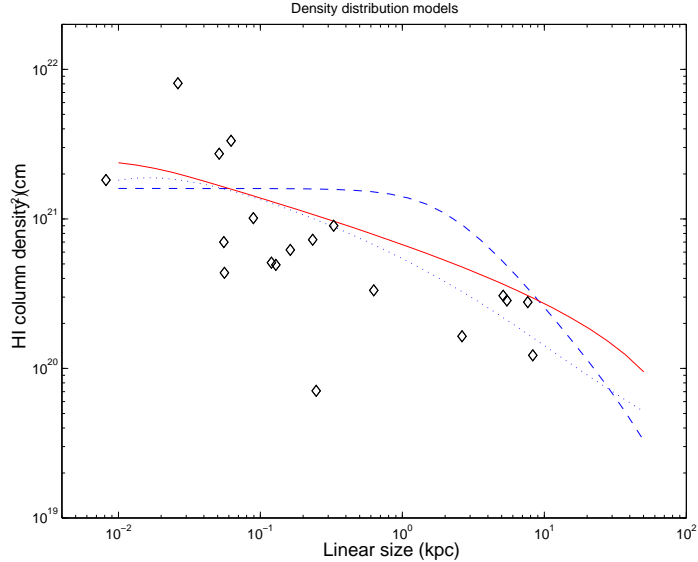,height=7.5cm}
\caption{Results of calculating the expected \hi\ column density given
different density profiles. The diamonds represents the real data. The
spherical King density profile, plotted with a dashed line, cannot
reproduce the observed \hi\ column density distribution for any value
of $\beta$. Instead, a power law drop of the density provides a more
similar distribution, with the closest fit for $\beta$=~1.25 and
$n_{\rm 0}=$~2.7\xxp{-2} \cmc\ (solid line). The dotted line
represents a disk distribution; it is possible to find reasonable fits
for a large range of opening angles.}
\label{models}
\end{center}
\end{figure}

\section{Origin of absorbing gas}
Our first results show that the column density decreases when the
linear size increases, implying a gas density which decreases with
radius. Such a density distribution could be either spherical or disk
like. By integrating the column density along the line-of-sight toward
the two lobes, we have investigated some functional forms for the
density distribution which may reproduce the observed correlation
between linear size and \hi\, optical depth.  We have assumed a
representative viewing angle of 45deg. Figure 3 shows a spherical and
an axi-symmetric (disk-like) distribution, both with a simple power
law radial fall-off in density, $n=n_{\rm 0}(r/r_{\rm 0})^{\rm
-\beta}$ ($r_{\rm 0}=1$ kpc) and also a King density profile with a
cutoff outside 50 kpc $n=n_{\rm 0}$ (1+$r^{\rm 2}/r_{\rm c}^{\rm 2}$)
($r_{\rm c}=1$ kpc). While King profiles cannot provide a good fit, in
contrast the power law models can fit the data adequately both for
spherical and disk-like geometries. For more details, see Pihlstr\"om
et al (2002, in prep.).

In most cases the spatial resolution of present \hi\ data is not
enough to determine whether the absorption covers both lobes
(consistent with ISM gas) or only one lobe (indicating a disk
distribution). To date high resolution \hi\ data exist for a limited
number of sources that in all cases indicate \hi\ disks (Conway 1996;
Peck et al.\ 1999; Peck \& Taylor 1998), which argues for a disk
model. One possibility to distinguish between a disk or a sphere is to
look at any possible orientation effects; in general quasars are
supposed to be seen at a smaller viewing angle than the radio
galaxies. Because of relativistic boosting the core strength is
considered to be a good indicator of orientation; Saikia et al.\
(1995) studied a sample of CSS sources with detected radio cores and
found that the degree of core prominence were consistent with those
for larger radio sources and also consistent with the unified
scheme. Assuming the unified scheme holds for GPS/CSSs, a spherically
symmetric distribution could thus be considered less likely, since we
in fact have found that there appears to be a larger detection rate in
galaxies ($\sim 60\%$) than in quasars ($\sim 25\%$; Pihlstr\"om
2001). We note however that there are indications that the GPS quasars
may not be the same type of objects as the GPS galaxies, in which case
we cannot use the argument of core prominence to distinguish between a disk or a
spherical origin of the gas. Instead, future higher resolution VLBI
observations (using the EVN UHF system) will be needed to determine
whether disks are the most common cause of the \hi\ absorption also in
the higher redshift objects.

\section*{Acknowledgments}
The WSRT GPS/CSS HI surveys are done in collaboration with:
P.D. Barthel, S.A. Baum, R. Braun, M.N. Bremer, W.H. de Vries,
G.K. Miley, C.P. O'Dea, H.J.A. R\"ottgering, R.T. Schilizzi,
I.A.G. Snellen, G.B. Taylor and W. Tschager.

\section*{References}





\reference Conway, J. E. 1996, in 'The Second Workshop on Gigahertz
Peaked Spectrum and Compact Steep Spectrum Radio Sources', eds. I.\
Snellen, R. T.\ Schilizzi, H. A. J.\ R\"ottgering \and M. N.\ Bremer, Publ
JIVE, Leiden, 198

\reference de Vries, W. H., O'Dea, C. P., Barthel, P. D., Fanti, C.,
Fanti, R. \and Lehnert, M. D. 2000, AJ, 120, 2300


\reference Gelderman, R. \and Whittle, M. 1994, ApJS, 91, 491

\reference Kameno, S., Horiuchi, S., Shen Z.-Q., Inoue, M., Kobayashi,
H., Hirabayashi, H. \and Murata, Y., 2000, PASJ 52, 209


\reference Kellermann, K. I., Vermeulen, R. C., Cohen, M. H., Zensus,
J. A. 1999, BAAS, 31, 856

\reference Lavalley, M., Isobe, T. \and Feigelson, E. 1992, BAAS, 24, 839

\reference Marr, J. M., Taylor, G. B. \and Crawford, F. III 2001, ApJ, 550, 160

\reference O'Dea, C. P., Stanghellini, C., Baum, S. \and Charlot,
S. 1996, ApJ, 470, 806

\reference Peck, A. B. \and Taylor, G. B. 1998, ApJ, 502, L23

\reference Peck, A. B., Taylor, G. B. \and Conway, J. E. 1999, ApJ, 521, 103

\reference Perlman, E. S., Stocke, J. T., Conway, J. E. \and Reynolds,
C. 2001, AJ, 122, 536

\reference Pihlstr\"om, Y. M. 2001, PhD Thesis, Chalmers University of
Technology, G\"oteborg

\reference Saikia, D. J., Jeyakumar, S. , Wiita, P. J., Sanghera,
H. S. \and Spencer, R. E.  1995, A\&A 295, 629

\reference Stanghellini, C., O'Dea, C. P., Dallacasa, D., Baum, S. A.,
Fanti, R. \and Fanti, C. 1998, A\&AS, 131, 303

\reference Vermeulen, R. C., Ros, E., Kellermann, K. I., Cohen, M. H.,
Zensus, J. A., van Langevelde, H. J. 2002a, PASA in press 

\reference Vermeulen, R. C., Ros, E., Kellermann, K. I., Cohen, M. H.,
Zensus, J. A., van Langevelde, H. J. 2002b, Submitted to A\&A

\end{document}